\documentclass[prb,twocolumn,balancelastpage,letterpaper,groupedaddress]{revtex4-1}
\usepackage{times}

\usepackage{dsfont}
\usepackage{amsmath}
\usepackage{amssymb}
\usepackage{graphicx}
\usepackage{bm}
\usepackage{bbm}
\usepackage{color}
\usepackage{verbatim}
\usepackage{tikz}
\usepackage{hyperref}
\usepackage{float}

\newcommand	{\ve}		{\varepsilon}

\begin{document}

\title{Dynamical competition between Quantum Hall and Quantum Spin Hall effects}

\author{A. Quelle and C. \surname{Morais Smith}}
\affiliation{Institute for Theoretical Physics, Center for Extreme Matter and Emergent Phenomena, Utrecht University, Leuvenlaan 4, 3584 CE Utrecht, The Netherlands}

\date{\today}
\pacs{00.00.-x}

\begin{abstract}
In this paper, we investigate the occurrence of quantum phase transitions in topological systems out of equilibrium. More specifically, we consider graphene with a sizable spin-orbit coupling, irradiated by circularly-polarized light. In the absence of light, the spin-orbit coupling drives a quantum spin Hall phase where edge currents with opposite spins counter-propagate. On the other hand, the light generates a time-dependent vector potential, which leads to a hopping parameter with staggered time-dependent phases around the benzene ring. The model is a dynamical version of the Haldane model, which considers a static staggered flux with zero total flux through each plaquette. Since the light breaks time-reversal symmetry, a quantum Hall (QH) phase protected by an integer topological invariant arises. An important difference with the static QH phase is the existence of counter-propagating edge states at different momenta, which are made possible by zero- and two-photon resonances. By numerically solving the complete problem, with spin-orbit coupling and light, and investigating different values of the driving frequency $\omega$, we show that the spectrum exhibits non-trivial gaps not only at zero energy, but also at $\omega/2$. This additional gap is created by photon resonances between the valence and conduction band of graphene, and the symmetry of the spectrum forces it to lie at $\omega/2$. By increasing the intensity of the irradiation, the topological state in the zero energy gap undergoes a dynamical phase transition from a quantum spin Hall to a quantum Hall phase, whereas the gap around $\omega/2$ remains in the quantum Hall regime.
\end{abstract}

\maketitle


\section{Introduction}
Topological phases of matter are insulating in the bulk, but feature metallic states at the boundaries. In the band structure of the system, these boundary states cross the gap between two bulk energy bands. The order parameter is a topological invariant characterizing the electronic band structure.\cite{Kitaev2009} This topological invariant is quantized, and insensitive to small perturbations of the system. The possible topological phases that a system can exhibit are strongly dependent on the symmetries which are broken or preserved.\cite{Ryu2010, Chiu2013} Indeed, breaking a symmetry removes the protection of the associated topological phases. An example is the quantum spin Hall (QSH) phase for 2D systems,\cite{Kane2005Z2,Kane2005QSHE,Zhang2010} which is protected by time-reversal (TR) symmetry. In the QSH phase, a material hosts counter-propagating edge states of opposite spin that are conjugated by TR. A single TR invariant pair (TRIP) of edge states cannot self-annihilate due to its topological protection. However, a pair of TRIPs can, causing the parity (an element of $\mathds{Z}_2$) of the number of TRIPs to be a topologically protected quantity.\cite{Kane2005Z2,Kane2005QSHE} This is in contrast to the quantum Hall (QH) state, which occurs when the TR symmetry is broken,\cite{Haldane1988} as happens, for instance, when a magnetic field is applied to the system. In the QH phase, gapless edge states also exist, except that they are unrelated by TR symmetry; the topological invariant can take any value in $\mathds{Z}$, and yields the number of edge states.\cite{Thouless1982}

Graphene was originally predicted to host a QSH effect due to the presence of intrinsic spin-orbit (ISO) coupling,\cite{Kane2005Z2,Kane2005QSHE} but it was soon discovered that the ISO coupling was two orders of magnitude too small to be of any notable effect.\cite{Min2006} However, recent results point out that heavily curved graphene samples, such as carbon nanotubes, do exhibit appreciable ISO coupling.\cite{Steele2013} In addition, doping graphene with heavy adatoms, such as In and Tl, may also lead to strong ISO coupling.\cite{Weeks2011} These discoveries revive the possibility of observing a QSH effect in graphene. Aside from graphene, the QSH effect has been predicted and experimentally verified in various other systems, such as HgTe quantum wells,\cite{Bernevig2006, Konig2007, Hasan2010} Bi$_2$Se$_3$ and Bi$_2$Te$_3$ crystals,\cite{Zhang2010} amongst others. Furthermore, a recent experimental realisation of synthetic graphene with truncated semiconductor nanocrystals has increased the expectation to create the QSH effect in honeycomb lattices.\cite{Kalesaki2014}

The behaviour of topological phases under the breaking of a protecting symmetry has been an object of study over the past few years. An interesting outcome is the possibility to drive a phase transition without closing the gap.\cite{EzawaGap2013} Overall, the behaviour of topological phases in equilibrium is well understood by now. A typical example is provided by breaking TR symmetry in a non-trivial QSH insulator. When this symmetry is broken, the edge states lose their topological protection. However, this does not imply that they will immediately annihilate due to scattering effects; merely breaking TR does not necessarily create the scattering effects that lead to their disappearance.\cite{Yang2011,Goldman2012} When a system is in a non topologically protected QSH phase, one speaks of a weak QSH phase\cite{Goldman2012} or a TR symmetry broken QSH phase.\cite{Yang2011} One way of creating such a weak QSH phase is by taking graphene with ISO coupling and applying a perpendicular magnetic field.\cite{Yang2011,Goldman2012} In this case, a competition between the QH and QSH phases is observed, where the dominating effect depends on the strength of the magnetic field and the ISO coupling. For weak $\bm B$ fields the QSH effect prevails, but by increasing the magnetic field a transition to a QH phase is induced. Interestingly, what constitutes a weak $\bm B$ field depends strongly on the ISO coupling, and for certain parameters, a transition from a weak QSH phase to a QH phase can also be driven by increasing the ISO coupling strength.\cite{Beugeling2012}

In recent years, it has been found that topological behaviour is also possible in out-of-equilibrium systems, where the time dependence is periodic. The behaviour of these so-called Floquet topological insulators is more complicated than in the time-independent case, but in the absence of TR symmetry, where QH states appear, the problem has been thoroughly investigated.\cite{Kitagawa2010, Lindner2011, Katan2013, Rudner2013} Experimentally, Floquet states have been observed on the boundary of Bi$_2$Se$_3$ in a topological insulator phase, using ARPES.\cite{Wang2013} Recently, it has been proposed that the edge states should also be directly visible using an STM tip to probe a HgTe Floquet system.\cite{Fregoso2014} Furthermore, Floquet topological systems have been simulated in photonics.\cite{Rechtsman2013} In this case, a honeycomb lattice of spiraling waveguides has been constructed, where the length of the waveguides functions as an artificial time parameter. When the photons are injected along the length of the waveguides, they hop between different waveguides, and the hopping becomes position dependent due to the twist of the waveguides. In this way, a honeycomb lattice with a time-dependent hopping is simulated. The first theoretical examples of Floquet topological insulators\cite{Lindner2011, Katan2013} predate the experimental detection, and were obtained by perturbing the Bernevig-Hughes-Zhang (BHZ) model.\cite{Bernevig2006} Currently, however, irradiated honeycomb lattices, such as graphene,\cite{Inoue2010, Gu2011, Kitagawa2011, Gomez-Leon2014, Perez-Piskunow2014, Usaj2014} and silicene\cite{Ezawa2013} are more prominent. These structures are of special interest due to their 2D nature, and the appearance of Dirac cones in the dispersion.

For graphene in particular, a host of interesting results already exists. In Ref.~\onlinecite{Kitagawa2011}, extensions of the Landauer-Buttiker formalism to Floquet systems have been written down and used to calculate a quantised Hall conductance at zero Fermi-energy for off-resonant light. This Hall conductance corresponds to a gap opening at the Dirac points, the size of which has been derived in Refs.~\onlinecite{Calvo2011} and \onlinecite{Fregoso2013} up to second order in the radiation amplitude. In these works, it is also noted that gaps open due to photon resonances. These gaps were later shown to be topological, and to host edge states.\cite{Gomez-Leon2014} In Ref.~\onlinecite{Sentef2014}, the ARPES response for irradiated Dirac electrons has been calculated at different moments of the irradiation period. Finally, analytical approximations for the wavefunctions and dispersions of edge states in graphene have been obtained by treating Floquet theory perturbatively.\cite{Perez-Piskunow2014, Usaj2014} These authors also demonstrate an excellent agreement of their approximations with numerics. Nevertheless, the effect of symmetry breaking on Floquet topological phases has not been explored yet.

In this work, the effect of broken TR symmetry in the Floquet case is investigated theoretically by applying circularly-polarized light to graphene with ISO coupling. It is shown that the breaking of TR symmetry leads to a competition between the QSH and QH phases in the system. This behaviour occurs in the static case,\cite{Beugeling2012} but the dynamical nature of the competition leads to interesting additional effects, such as photon resonances and the absence of Landau levels. Due to the absence of Landau levels, the QH phases found here are, properly speaking, quantum anomalous Hall (QAH) phases.

The outline of the paper is as follows. In Sec.~II, Floquet systems are discussed. We provide the necessary definitions, and expound upon the way in which Floquet systems manifest topological behaviour. In Sec.~III, graphene irradiated by circularly-polarized light is discussed in the absence of ISO coupling. First we introduce the model; subsequently, a physical interpretation is given. By looking at the system semi-classically, it can be shown that under circularly-polarized light an electron feels an effective magnetic flux as it moves through the lattice. This makes the appearance of QH states intuitively clear. Sec.~IV contains the main results of this work: how the results from Sec.~III change under the inclusion of ISO interaction. A competition between QH and QSH states is demonstrated, with the dominant effect being determined by the relative strength of the ISO coupling and the light field. Our conclusions, together with an outlook, are presented in Sec.~V.

\section{Floquet theory}

To set up the framework for the model, we first present a review of time-dependent systems, together with the way in which they can exhibit topological behaviour.  Systems where the time dependence is periodic, i.e. the Hamiltonian obeys $H(t+T)=H(t)$,  are particularly tractable. In this case, Floquet theory can be applied to analyze the behaviour of the systems. If $T$ is the period of the Hamiltonian, the basis of Floquet theory is the equality $U(t+T,t'+T)=U(t,t')$ for the propagator,
\begin{equation*}
U(t,t')=\mathcal{T}\left(\exp\left[-i\int_{t'}^t H(\tau)d\tau \right] \right),
\end{equation*}
where $\mathcal{T}$ denotes the time-ordering operator. This has the immediate consequence that the eigenfunctions of $U(T,0)$ (the Floquet states) are quasi-periodic in time with period $T$, and have a corresponding complex eigenvalue of unit norm. Physically, these eigenvectors are the analogues of the stationary states of the system; they will have a time dependence, but due to their quasi-periodicity, this time dependence is well controlled. In the high-frequency limit, where the period of the system is too short to be accurately measured, the Floquet states are stationary for all practical purposes. In this limit, the photon energy will be large compared to the bandwidth of the system, ruling out the possibility of photon resonances. We show below that for intermediate frequencies, these photon resonances may lead to an interesting topological phase transition.

As a rule, it is easier to analyze Hermitian operators than unitary ones, which leads to the definition of the Floquet Hamiltonian
\begin{equation}\label{floqham}
H_F:=-\frac{i}{T}\ln\left[U(T,0)\right].
\end{equation}
We are considering $\hbar=1$ throughout this work. The Floquet Hamiltonian has the same eigenfunctions as $U(T,0)$, but its eigenvalues are real; physically, $H_F$ is the time-averaged Hamiltonian for the system. The eigenvalues of $U(T,0)$ lie on the unit circle; to capture this behaviour we take the logarithm to be multivalued, rather than choosing a branch cut. The Floquet Hamiltonian will then have a periodic spectrum with period $\omega=2\pi/T$. Viewing the theory in terms of $H_F$ also emphasizes the interpretation of its eigenfunctions as the stationary states of the system. In the high-frequency limit, this point of view has added poignancy, because then the rapid fluctuations of the eigenfunctions belonging to $H_F$ become difficult to measure. In this case, one may average out the fluctuations and approximate the system by a stationary one, which is obviously $H_F$, due to its time-averaged nature.

It is of particular interest that the spectrum of $H_F$ can host gapless edge states, in which case we will speak of Floquet topological insulators.\cite{Lindner2011, Katan2013} To analyze the topological behaviour of Floquet systems, it is possible to calculate the relevant topological invariants of the Floquet bands (the bands of $H_F$).\cite{Fukui2005, Kitagawa2010, Rudner2013} However, the spectrum of $H_F$ is periodic, so there is no lowest band, and the bulk-boundary correspondence does not give complete information about the system.\cite{Kitagawa2010, Rudner2013} Rudner~\textit{et~al.}\cite{Rudner2013} have succeeded in deriving a topological invariant which accurately yields the chirality of each gap. This invariant measures the charge that the states traversing a given gap pump around the system (if it is cylindrical), or across it (if it is a ribbon), in a single period of the Hamiltonian. Since the number of edge states traversing each gap is integer, the same holds for the invariant. Mathematically, this follows because the invariant is a winding number. Therefore, in the absence of symmetry, the chiralities of each gap are the topological invariants of the system, analogous to the stationary case. Furthermore, these chiralities can take any value in $\mathds{Z}$, indicating that we are dealing with a Floquet QH phase.

\section{Irradiated graphene}
Irradiating 2D materials with circularly-polarized light creates Floquet QH insulators in a variety of cases.\cite{Inoue2010, Gu2011, Ezawa2013, Wang2013, Gomez-Leon2014} Let us focus on graphene, for which the nearest-neighbor (nn) tight-binding Hamiltonian reads\cite{CastroNeto2009}
\begin{equation}\label{Hnaught}
H_0=-J \sum_{\langle i,j \rangle} c^\dagger_i c_j.
\end{equation}
Here, $J\geq 0$ is the nn hopping strength, which is usually denoted by $t$, but is renamed due to the use of $t$ as the time parameter; the brackets denote a sum over nn only. It is important to note that $H_0$ is still symmetric under TR. We modulate $H_0$ by irradiating the graphene sample with circularly-polarized light. The light can be described by a vector potential, and incorporated into the Hamiltonian through the Peierls substitution. The vector potential is taken to be
\begin{equation}\label{vecpot}
\bm A(t)=V \left(\cos(\omega t),\sin(\omega t)\right).
\end{equation}
The Peierls substitution consists of sending $J\mapsto J_{i,j}$, where
\begin{equation}\label{Peierls}
J_{i,j}(t)=J \exp\left[-i e \int_i^j \bm A(t)\cdot d\bm s\right].
\end{equation}
Here, the path from site $i$ to $j$ is irrelevant, since there is no perpendicular component to the $\bm B$ field. The final Hamiltonian for the model reads
\begin{equation}\label{Hamiltonian}
H(t)=-\sum_{\langle i,j \rangle}J_{i,j}(t) c^\dagger_i c_j,
\end{equation}
where $J_{i,j}$ is given by Eqs.~(\ref{vecpot}) and~(\ref{Peierls}). This full Hamiltonian $H$ is no longer TR symmetric; this occurs because the rotational direction of the light changes under TR. Before the results for this model are given, a physical interpretation for the phases in $J_{i,j}$ is provided. From this reasoning, it should be intuitively clear that a QHE is generated, as confirmed by calculations.
\subsection{The physical interpretation: a dynamical Haldane model}
\begin{figure}[h]
\includegraphics{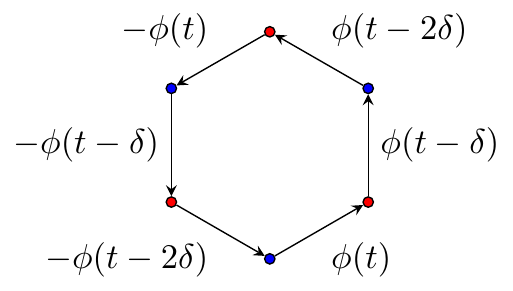}
\caption{\label{hopping phases}A plaquette of the honeycomb lattice is shown. Sites of the A (B) sub lattice are depicted as red (blue) dots. The arrows denote counter-clockwise hopping along the lattice bonds by an electron. Next to each arrow the hopping phase $\phi$ is given. Important is that all the hopping phases have the same functional form, but due to a different orientation of the bonds, they are evaluated at times differing by a multiple $\delta=T/6$.}
\end{figure}

In Fig.~\ref{hopping phases}, the hopping of an electron along a plaquette of the honeycomb lattice is shown. The electron picks up a complex phase with each hopping due to the presence of a vector potential. All these phases have the same form $\phi(t)$, except that they are evaluated at different times due to the different orientations of the bonds and the rotation of the vector potential in time. By varying the electron position, the hopping phase, and hence the vector potential, is probed at a different time. The vector potential rotates in time, with the time averaged vector potential being zero. As a result, an effectively rotating vector potential is seen in space, which creates a magnetic field. Because the time-averaged vector potential vanishes, the effective net magnetic flux also vanishes. 

The situation can be interpreted as a dynamical analogue of the Haldane model.\cite{Haldane1988} In Haldane's model, the honeycomb lattice is subjected to staggered magnetic fluxes such that the electron picks up a phase while hopping between next-nearest neighbors (nnn). By varying the position of the electron, a varying vector potential is felt which corresponds to a magnetic field with zero net flux. Consequently, a QAH effect is created; QH edge states are observed, but the usual splitting in Landau levels does not occur because of the vanishing net magnetic flux. In the case of circularly-polarized light, the electrons pick up a phase when hopping between nn, while there is no net flux through the hexagonal plaquettes. This causes the appearance of QH edge states, without leading to the occurrence of Landau levels.\cite{Gomez-Leon2014}
\subsection{The band structure of irradiated graphene}
\begin{figure*}
\includegraphics[width=\textwidth]{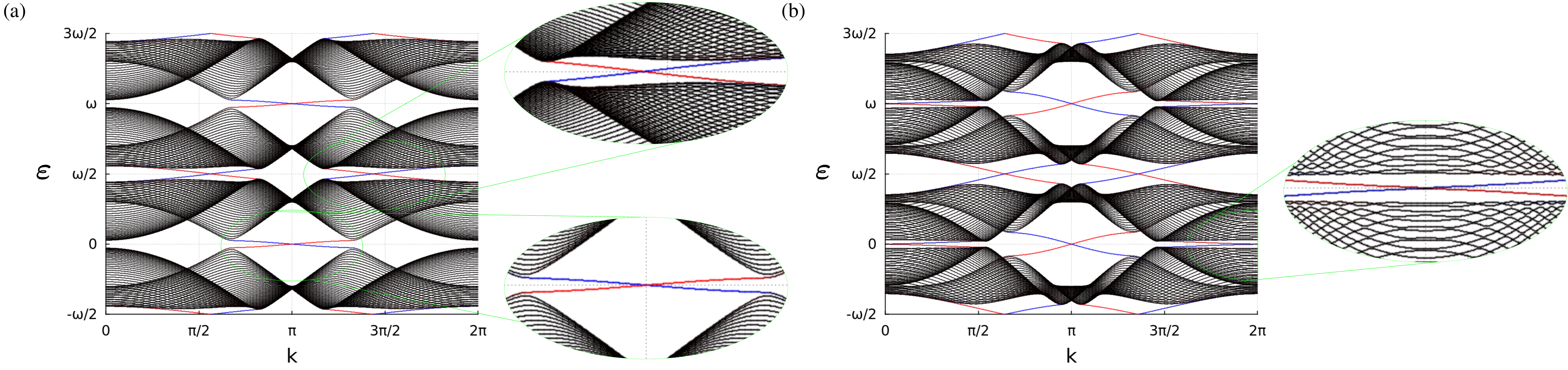}
\caption{\label{graphs}(Color online) The spectrum of the Floquet Hamiltonian defined in Eqs.~(\ref{floqham}) and (\ref{Hamiltonian}), for a graphene cylinder with zigzag edges; note that the entire spectrum is spin degenerate. Even though there are only two energy bands in graphene, the periodicity of the spectrum causes the existence of two inequivalent band gaps. The inlays show the gapless edge states with a magnification of two for greater clarity. Black bands are localised in the bulk, and red/blue bands are localised on the upper/lower edge of the system. Parameter values are (a) $V=J,\ \omega=3J$ and (b) $V=J,\ \omega=2.2J$. Since the band gaps open at multiples of $\omega/2$ for all values of $\omega$, the quasi-energies $\varepsilon$ are given in units of $\omega/2$, which takes a different value in (a) and (b).}
\end{figure*}
By analyzing the system semi-classically, it becomes intuitively clear that circularly-polarized light creates a QH effect. The precise manner in which this happens is depicted in Figs.~\ref{graphs}(a) and \ref{graphs}(b) for a graphene cylinder with zigzag edges. Here, the spectrum for the Floquet Hamiltonian defined through Eqs.~(\ref{floqham}) and (\ref{Hamiltonian}) is shown for two different values of light frequency: $\omega=3 J$ in Fig.~\ref{graphs}(a) and $\omega=2.2 J$ in Fig.~\ref{graphs}(b). In the regime where $\omega$ is large, analytical approximations for $H_F$ exist,\cite{Eckardt2005,Koghee2012} which take the form of a static tight-binding Hamiltonian with some renormalised coefficients. For lower frequencies, one can use an extension of this same method, but one has to take into account the periodicity of $H_F$.\cite{Usaj2014} This method is also used to derive analytical expressions for gap sizes\cite{Calvo2011, Fregoso2013}, and conductances.\cite{Kitagawa2011} The calculations performed in Refs.~\onlinecite{Calvo2011, Kitagawa2011, Fregoso2013} were done for an infinite plane, whereas we work on a ribbon because we are interested in obtaining the edge states. Consequently, we found it more convenient to use a different approach. By numerically solving the Schr\"odinger equation corresponding to the Hamiltonian in Eq.~(\ref{Hamiltonian}), a numerical approximation to $U(T,0)$ was acquired. This in turn was used to obtain a numerical approximation to $H_F$ through the use of Eq.~(\ref{floqham}). It should be noted that Fig.~\ref{graphs}(a) gives similar results to those obtained in Ref.~\onlinecite{Usaj2014}. In Ref.~\onlinecite{Usaj2014}, the spectrum of a graphene half-plane is calculated for the same photon energy, but lower intensity, by the method of recursive Green's functions. 

In Fig.~\ref{graphs}, two inequivalent gaps are seen, each repeating with period $\omega$: one at $\varepsilon=0$ and one at $\varepsilon=\omega/2$. First, we will consider the spectrum in Fig.~\ref{graphs}(a). In the gap around $\ve=0$, a single spin degenerate edge state is observed. It should be noted that the entire spectrum is spin degenerate due to the absence of Zeeman splitting in Eq.~(\ref{Hamiltonian}). The light opens a gap at the Dirac points, imparting an effective mass to the low-energy Floquet states. The opened gap is topological, hosting QH states, since the electrons feel an effective magnetic field. This behaviour is characteristic of the QAH effect, since no Landau levels are created. By inspection, the chirality of this gap is unity for each spin; for each spin a single gapless state exists on both edges. This gap also exists in the absence of photon resonances, since it is caused by the effective magnetic field experienced by the electrons. In the off resonant regime, it will lead to the quantised Hall conductance calculated in Ref.~\onlinecite{Kitagawa2011}. Up to third order, the size of the gap at the Dirac points is quadratic in $V$, and it exists for all light intensities.\cite{Calvo2011, Fregoso2013} On the other hand, the gap around $\ve=\omega/2$ has a chirality of two for each spin. The origin of these edge states is different, however. Because of the $\omega$-periodicity of the Floquet spectrum, the conduction band above the $\ve=0$ gap overlaps with the valence band below the $\ve=\omega$ gap, i.e. the valence band one period higher. The symmetry of the graphene spectrum in $\ve\mapsto -\ve$ imposes that this happens symmetrically around $\ve=\omega/2$. At this point, the energy difference between valence and conduction band is precisely $\omega$, allowing for photon resonances. These photon resonances cause hybridization of states in the valence and conduction bands, opening a gap. Its size is linear in $V$ up to third order in this quantity.\cite{Calvo2011, Fregoso2013} This gap also exists for all intensities, and the different scaling with light intensity originates in the different nature of the gap. Because the electrons still feel an effective magnetic field, this gap is also topological, hosting QH states.

In Fig.~\ref{graphs}(b), one may observe the same edge states as in Fig.~\ref{graphs}(a), with the addition of an extra edge state in the $\ve=0$ gap. The edge states in the $\ve=\omega/2$ gap and those crossing at $k=\pi$ in the $\ve=0$ gap have the same origin as those in the case where $\omega=3J$. The extra edge state crossing at $k=0$ (or $k=2\pi$) in the $\ve=0$ gap (see the inlay of Fig.~\ref{graphs}(b)) is also caused by photon resonances. Because $\omega$ (and hence the periodicity) has decreased, the conduction band above the gap at $\ve=-\omega$ has started to overlap with the valence band below the gap at $\ve=\omega$. By symmetry, the overlap has to happen around $\ve=0$, and at this point the energy difference between valence and conduction band is $2\omega$. In this case, two photon absorption creates resonances between the valence and the conduction bands, once again opening a topological gap. It should be noted that the two pairs of edge states counter-propagate; the edge state caused by the two-photon resonances propagates oppositely to the edge state at the Dirac points, which is not caused by photon resonances. Similar behaviour has been observed in the driven Hofstadter model.\cite{Lababidi2014, Zhou2014} In the Hofstadter model, the edge states appearing in the gap at $\varepsilon=\omega/2$ always appear in counter-propagating pairs, making this gap trivial. In Refs.~\onlinecite{Lababidi2014} and \onlinecite{Zhou2014}, the stability of these counter propagating states with respect to disorder has been investigated. Although disorder couples different crystal momenta, this does not cause the edge states to annihilate. Even though the gap is trivial, the counter-propagating states are relatively stable against perturbations of the Hamiltonian. Because the width of the valence and the conduction band in bare graphene [given by Eq.~(\ref{Hnaught})] is $3J$ (and hence, the total width is $6J$), resonances at $\omega$ and $2\omega$ are possible for $\omega=2.2J$. Although one would expect that for $\omega=3J$ two-photon resonances could still appear, in reality they do not because the light causes a slight flattening of the energy bands, thus decreasing the bandwidth. This is evident from Fig.~\ref{graphs}(a), where the bands do not meet at $\ve=0,\ k=0$.
\begin{figure*}[t]
\includegraphics[width=\textwidth]{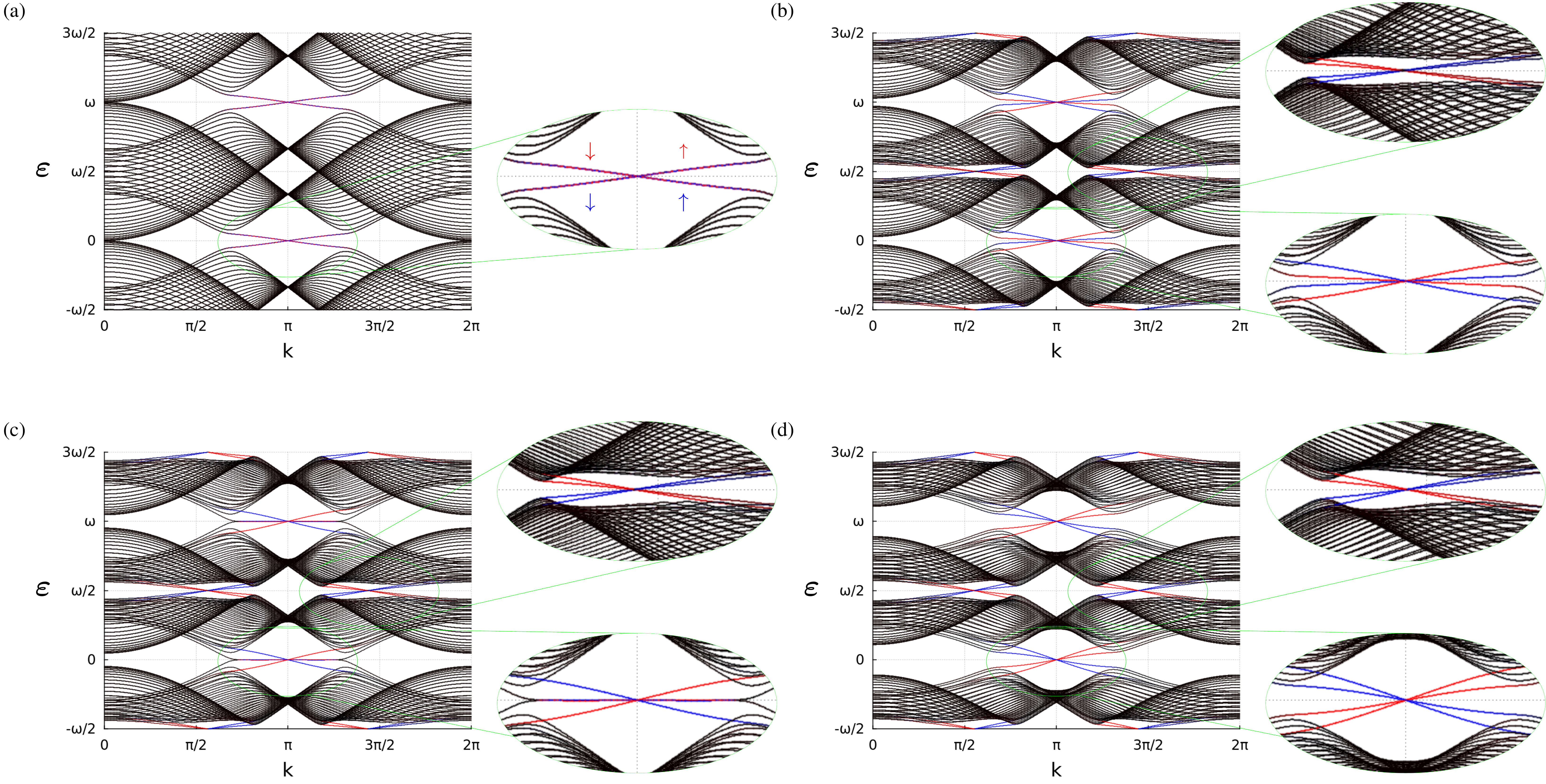}
\caption{\label{graphsISO}(Color online) The spectrum of the Floquet Hamiltonian defined in Eqs.~(\ref{floqham}) and (\ref{FloqHamISO}) for $\omega=3J$ and $\lambda=0.06J$, for a graphene cylinder with zigzag edges. The inlays show the gapless edge states with a magnification of two, for greater clarity. Black bands are localised in the bulk and red/blue bands are localised on the upper/lower edge of the system. (a) For $V=0$, the known dispersion for the QSH effect is observed, apart from a periodic continuation in $\omega$, caused by artificially considering the Floquet spectrum despite a time independence. Since the spectrum is spin degenerate, we indicated the spin direction for each edge state in the inlay. (b) For $V=J$, in the presence of both ISO coupling and circularly-polarized light, the spin degeneracy in both edge and bulk states is lifted. In the gap at $\ve=0$, the QSH still dominates. (c) For $V=1.3J$ and $\ve=0$, the QH and QSH effects annihilate each other for spin down, and enhance each other for spin up. This causes one of the states on each edge to become dispersionless, and thus, also localised. (d) For V=$2J$, the irradiation dominates; aside from a lifting of the spin degeneracy, the graph resembles the one in Fig.~\ref{graphs}(a), where QH states are generated.}
\end{figure*}

\section{The interaction between spin-orbit coupling and light}
It has been demonstrated by Steele et~al.\cite{Steele2013} that in heavily curved graphene surfaces (for example carbon nanotubes) an appreciable ISO coupling is present. It is well known\cite{Kane2005Z2,Kane2005QSHE} that the ISO coupling opens a topological gap in graphene, hosting a TRIP of edge states. Mathematically, the two bands of graphene have a non-vanishing $Z_2$ quantum number. In the QSH effect, the propagation direction of the edge states is spin-polarized, whereas in the QH effect it is not. This leads to a competition between the two effects, as the QH effect becomes more pronounced and TR symmetry is increasingly violated. 
\subsection{The model}
To model this competition, we consider the tight-binding Hamiltonian\cite{Kane2005QSHE,Kane2005Z2}
\begin{equation}\label{FloqHamISO}
H(t)=-\sum_{\langle i,j \rangle}J_{i,j}(t) c^\dagger_i c_j-i\sum_{\langle\langle i,j \rangle\rangle}\hat{s}_z\lambda_{i,j}(t)\nu_{i,j}c^\dagger_i c_j,
\end{equation}
where double brackets denote a sum over nnn. Furthermore, $J_{i,j}$ is given by Eq.~(\ref{Peierls}), and 
\begin{equation}\label{PeierlsISO}
\lambda_{i,j}(t)=\lambda \exp\left[-i e \int_i^j \bm A(t)\cdot d\bm s\right],
\end{equation}
with $\lambda$ denoting the ISO coupling strength. Additionally, $\nu_{i,j}=\pm 1$ and $\hat{s}_z$ is the spin operator, yielding the sign change of the ISO interaction for the different spins. Note that $\hat s_z=\hat \sigma_z/2$, where $\sigma$ is a Pauli matrix. This is similar to the ISO Hamiltonian from Kane \& Mele.\cite{Kane2005QSHE,Kane2005Z2} Importantly, the hopping is now modulated by the vector potential that incorporates the light field; this causes both the nn and the nnn hopping to have a real and an imaginary component. In the Kane \& Mele model, the nn hopping is real, and the nnn hopping is purely imaginary. Apart from a difference in vector potential, this is the model considered by Beugeling~\textit{et~al.}\cite{Beugeling2012} In Ref.~\onlinecite{Beugeling2012}, the authors consider a perpendicular magnetic field and $\nabla\times \bm A=\bm B$. Here, according to Eq.~(\ref{vecpot}), $\nabla\times \bm A=0$. The non-trivial topological phases in this case come from the time dependence of $\bm A$; it is this dynamical character that leads to the unexpected behaviour.

\subsection{The results}

The dispersion relation for the Floquet Hamiltonian defined in Eqs.~(\ref{floqham}) and (\ref{FloqHamISO}) is shown in Fig.~\ref{graphsISO} for various values of the relevant parameters. The system has a cylindrical geometry with zigzag edges, and the spectra were obtained using the same method as without SO coupling. 

In Fig.~\ref{graphsISO}(a), the spectrum of the system is shown in the absence of light. In this case, the known dispersion is recovered, apart from a periodicity in $\omega$, which is manually inserted to facilitate comparison with the time-dependent cases. In Fig.~\ref{graphsISO}(b), all parameters are the same as for the dispersion in Fig.~\ref{graphs}(a), apart from the addition of the ISO coupling. Here, the light field opens a gap at $\omega/2$, which hosts the same type of edge states as in the absence of ISO coupling. The ISO coupling cannot have an impact on the topological nature of this gap because it is caused by photon resonances between the conduction and valence bands; the ISO coupling cannot create such resonances. The same occurs for the two photon resonances at $\ve=0$ for $\omega=2.2J$ (not shown): the degeneracy of the resulting edge states is lifted, but they will be QH states regardless of light intensity. In the gap at $\ve=0$, edge states of a QSH type are observed. Although the light breaks the TR symmetry of the system, and the QSH states are no longer protected, the perturbing effect of the light is not enough to break the QSH-like nature of the edge states. In both gaps, the degeneracy of the edge states is lifted and two states of different velocity are created on each edge. The slope of the state is related to the spin orientation: the slope is increased for spin up, where the QH and QSH effect enhance each other, whereas for spin down they interfere destructively and the slope is decreased. Furthermore, the degeneracy of the bulk states is also lifted. The ISO coupling alone cannot do this, but when it is combined with an applied magnetic field\cite{Beugeling2012}, or with circularly-polarized light, this degeneracy is lifted. It is also possible to lift the degeneracy by inducing a Rashba spin-orbit coupling.\cite{Zarea2009, Gelderen2010}
In Fig.~\ref{graphsISO}(c), the intensity of the light is further increased. As before, the nature of the edge states in the gap at $\ve=\omega/2$ is unchanged, whereas the edge states in the gap at $\ve=0$ are affected by the increased competition between the QH and QSH effect. For the spin value where they enhance each other, the velocity of the edge state is increased, while for the other spin the two effects annihilate each other. The gap closes, the edge state becomes dispersionless, and it becomes localised.
In Fig.~\ref{graphsISO}(d), the light intensity has become so high that the QH effect also dominates in the gap at $\ve=0$. The ISO coupling still causes a lifting of the spin degeneracy of the edge states, but all the gapless states are of a QH nature (see Fig.~\ref{graphs}(a) for a comparison). Specifically, the gap closing at $\ve=0$ has led to a change in Chern number of two (since one of the spins changes propagation direction) for both bands, corresponding to the change in chirality of the gap. 

These are the main results of this paper: The combination of ISO coupling and circularly-polarized light lifts the spin degeneracy of both the edge and the bulk states. In addition, a dynamical quantum phase transition occurs between a QSH and a QH state in the $\ve=0$ gap upon increasing the intensity of the circularly-polarized light. Furthermore, at the topological quantum phase transition, one spin is localised, whereas the other spin yields a quantised edge current. The parameter values in this work have been chosen to draw plots in which the relevant topological features are easily distinguished. Similar parameters have also been used previously,\cite{Usaj2014} which allows for comparison of the results. To investigate the phase transition point between QH and QSH effect, observe that the gap $\Delta_r$ opened by the radiation at the Dirac points reads\cite{Calvo2011, Fregoso2013}
\begin{equation*}
\Delta_r=\frac{J^2e^2a^2V^2}{\omega}+J\mathcal{O}(eaV)^3,
\end{equation*}
where $a=0.246$nm is the graphene lattice constant. This shows that for ISO induced gaps $\Delta_{\rm{ISO}}$ that are small with respect to $J\simeq 2.8$eV,\cite{CastroNeto2009} one finds a phase transition at
\begin{equation*}
E^2=\frac{\omega^3\Delta_{\rm{ISO}}}{J^2e^2a^2},
\end{equation*}
where $E=V/\omega$ is the electric field strength. Assuming a photon frequency $\omega\simeq100$meV, and that the ISO coupling opens a gap $\Delta_{\rm{ISO}}\simeq3$meV,\cite{Steele2013} yields a field strength $E\simeq10^6$V/m, which is achievable experimentally. This indicates that the described phase transition is accessible in the laboratory.

\section{Conclusions}\label{sect_conclusion}
The topological behaviour of an insulator is strongly dependent on its symmetries. By breaking a symmetry of the system, edge states can lose their topological protection. In this situation, such a phase can be gapped out. After the gap closes and reopens, a new topological phase may appear, which was not possible before the symmetry was broken. An example of this is found in the competition between the QH and QSH effects in graphene,\cite{Yang2011,Goldman2012} where a strong magnetic field destroys the QSH effect. A QH effect, characteristic of perpendicular magnetic fields applied to 2D systems, appears instead. A competition between these two effects can also be seen in the Floquet case if the TR symmetry is broken by circularly-polarized light. The Floquet spectrum of the system will then exhibit a QSH effect at the $\varepsilon=0$ gap for weak light intensities, and for increasing intensities it will undergo a transition to a QH effect. This QH effect appears because the rotating vector potential creates an effective magnetic field for the electrons moving through the lattice. The various QH-like edge states created by photon resonances are not affected by the ISO coupling, because this coupling cannot create such resonances. Consequently, it cannot interfere with such resonant states either. The possibility of photon resonances allows for a dynamical tuning of Floquet systems that is not available in the static case. Through this example, it becomes clear that symmetry breaking in the time-dependent case leads to richer behaviour than is possible in  equilibrium systems. Additionally, this behaviour is predicted to be experimentally accessible. We hope that our work will motivate experiments in irradiated single-layer graphene doped with heavy ion adatoms, as well as in carbon nanotubes.

\acknowledgments
We thank V.\ Juri\v{c}i\'c for useful discussions. The work by A.Q.\ is part of the D-ITP consortium, a program of the Netherlands Organisation for Scientific Research (NWO) that is funded by the Dutch Ministry of Education, Culture and Science (OCW). C.M.S.\ acknowledges NWO for funding within the framework of a VICI program.

\bibliographystyle{apsrev4-1}
\end{document}